# On the puzzling feature of the silence of precursory electromagnetic emissions.

K. Eftaxias[a], S. M. Potirakis[b], and T. Chelidze[c]

a. Department of Physics, Section of Solid State Physics, University of Athens, Panepistimiopolis, GR-15784, Zografos, Athens, Greece, ceftax@phys.uoa.gr .
b. Department of Electronics Engineering, Technological Education Institute (TEI) of Piraeus, 250 Thivon & P. Ralli, GR-12244, Aigaleo, Athens, Greece, spoti@teipir.gr .
c. M. Nodia Institute of Geophysics at the I. Javakhishvili Tbilisi State University, 1, Alexidze str. Tbilisi, 0171, Georgia, tamaz.chelidze@gmail.com

**Abstract**

It has been suggested that fracture-induced MHz-kHz electromagnetic (EM) emissions, which emerge from a few days up to a few hours before the main seismic shock occurrence permit a real-time monitoring of the damage process during the last stages of earthquake preparation, as it happens at the laboratory scale. Despite fairly abundant evidence, EM precursors have not been adequately accepted as credible physical phenomena. These negative views are enhanced by the fact that certain "puzzling features" are repetitively observed in candidate fracture-induced pre-seismic EM emissions. More precisely, EM silence in all frequency bands appears before the main seismic shock occurrence, as well as during the aftershock period. Actually, the view that "acceptance of "precursive" EM signals without convincing co-seismic signals should not be expected" seems to be reasonable. In this work we focus on this point. We examine whether the aforementioned features of EM silence are really puzzling ones or, instead, reflect well-documented characteristic features of the fracture process, in terms of: universal structural patterns of the fracture process, recent laboratory experiments, numerical and theoretical studies of fracture dynamics, critical phenomena, percolation theory, and micromechanics of granular materials. Our analysis shows that these features should not be considered puzzling.

# 1. Introduction

In recent years, the wind prevailing in the scientific community does not appear to be favorable for earthquake (EQ) prediction research, in particular for the research of short term prediction (Uyeda et al., 2009). Sometimes the arguments have been extended to the extreme claim that any precursory activity is impossible (Geller et al., 1997). Considering the difficulties associated with such factors as the highly complex nature and rarity of large EQs and subtleties of possible pre-seismic signatures, the present negative views are not groundless.

An EQ is a sudden mechanical failure in the Earth's crust, which has heterogeneous structures. The employment of basic principles of fracture mechanics is a challenging field for understanding the EQ preparation process. A vital problem in material science is the identification of precursors of macroscopic defects. Fracture-induced physical fields allow a real-time monitoring of damage evolution in materials during mechanical loading. More specifically, electromagnetic emissions (EME) and acoustic emissions (AE) in a wide frequency spectrum ranging from the kHz to the MHz bands are produced by opening cracks, which can be considered as the so-called precursors of general fracture. Recently, improvements in the *MHz-kHz EME* technique have permitted a real-time monitoring of the fracture process (Fukui et al., 2005; Kumar and Misra, 2007; Chauhan and Misra, 2008; Baddari et al., 1999; Baddari and Frolov, 2010; Baddari et al., 2011; Lacidogna et al., 2010; Schiavi et al., 2011; Carpinteri et al., 2012). However, the MHz–kHz EM precursors are detectable not only at the laboratory but also at the geological scale; a stressed rock behaves like a *stress-EM transducer* (Sadovski, 1982; Hayakawa and Fujinawa, 1994; Gokhberg, et al., 1995; Hayakawa, 1999; Hayakawa and Molchanov, 2002; Eftaxias et al., 2007, 2011; Eftaxias, 2012; Molchanov and Hayakawa, 2008; Hayakawa, 2009). The idea that the fracture-induced MHz-kHz EM fields should also permit the monitoring of the gradual damage of stressed materials in the Earth's crust, as happens in the laboratory experiments, in real-time and step-by-step, seems to be justified: the aspect of self-affine nature of faulting and fracture is well documented . An interesting experimental research program would be the parallel monitoring of the corresponding observable manifestations of both laboratory and geophysical scale fracture phenomena. Based on this idea we have installed a field experimental network using the same instrumentation as in laboratory experiments for the recording of geophysical scale EME. An exemplary telemetric station, as part of this network has been operating on Zante Island (Greece) since 1994, mainly aiming at the detection of kHz-MHz EM precursors. It has been installed in a carefully selected mountainous site in the southwest part of the island (37.76º N–20.76º E) providing low EM background noise. The complete measurement system is comprised of (i) six loop antennas detecting the three components (EW, NS, and vertical) of the variations of the magnetic field at 3 kHz and 10 kHz respectively; (ii) three vertical $\lambda/2$ electric dipole antennas detecting the electric field variations at 41, 54 and 135 MHz respectively; and (iii) other magnetic and electromagnetic sensors. All the time-series are sampled once per second, i.e., with a sampling frequency of 1 Hz.

"Are there credible EM EQ precursors?" This is a question debated in the science community (Eftaxias, 2012). Despite fairly abundant evidence, EM precursors have not been adequately accepted as real physical quantities (Morgounov and Shakhraman'yan, 1996; Uyeda et al., 2009; Hayakawa and Hobara, 2010). There may be legitimate reasons for the critical views concerning the reliability of EM precursors: The degree to which we can predict a phenomenon is often measured by how well we understand it. However, many questions about the fracture processes remain standing. Especially, many aspects of EQ generation remain yet to be understood. The thorough understanding of EM precursors in terms of physics is a path to achieve deeper knowledge of the last stages of the EQ preparation process, and thus a path to more credible short-term EQ prediction. A *seismic* shift in thinking towards basic science will lead to a renaissance of strict definitions and systematic experiments in the field of EQ prediction. Kossobokov (2006) emphasizes that "No scientific prediction is possible without exact definition of the anticipated phenomenon and the rules, which define clearly in advance of it whether the prediction is confirmed or not."

As noted above, there may be legitimate reasons for the critical views concerning the reliability of EM precursors. These negative views are supported by the fact that specific "puzzling features" are systematically observed in candidate pre-seismic EM emissions. Specifically: (i) *EM silence in all frequency bands appears before the main seismic shock occurrence; and (ii) EM silence is also observed during the aftershock period.*

In this work we focus on this point. We examine whether the aforementioned features of EM silence are really puzzling ones or, instead, reflect well-documented characteristic features of fracture process.

The main concept is to investigate whether laboratory experiments reproduce these "puzzling features." It was suggested early that *"the mechanism of EQs is apparently some sort of laboratory fracture process"* (Mogi, 1962a, b; Mogi., 1968, 1985; Ohnaka, 1983; Ohnaka and Mogi, 1982; Scholz, 1968, 1990; Scholz et al., 1973; Sobolev et al., 1995; Ponomarev et al., 1997; Ohnaka, M., and Shen, 1999; Sobolev and Ponomarev, 2003; Muto et al., 2007; Kuksenko et al., 1996, 2005, 2007, 2009; Lei and Satoh, 2007). Recently, the aspect of self-affine nature of faulting and fracture is widely documented from analyses of data from both field observations and laboratory experiments in the spatial, temporal, and energy domains (Mandelbrot, 1982; Mandelbrot et al., 1984; Zavyalov et al., 1988; Gabrielov et al., 1999; Kuksenko et al., 2007; Davidsen and Schuster, 2002; Lockner et al., 1992; Diodati et al., 1991; Kapiris et al., 2004). A number of studies indicate a similarity of the statistical properties of laboratory seismicity in terms of AE and EME on one hand, and seismicity on the geological scale on the other hand, in the spatial, temporal, and energy domains. Therefore, we really have the opportunity to ask the following important question: *"Can we identify any accordance between the lab and the field by means of the two above-mentioned puzzling features?"*

Of course, it is a risky practice to extend findings rooted in laboratory experiments to the geophysical scale. However, one cannot ignore the comparison between failure precursors at the laboratory and geophysical scales, especially taking into account the ubiquitous power law in the fracture process, i.e., self-similarity between fracture patterns on various scales.

The remainder of this article is organized as follows. Sec. 2 presents the up-to-date theoretical and laboratory background; this includes a brief description of the two-stage model for the EQ generation process and the percolation theory of fracture, upon which the presented analysis is based. Based on the evidence highlighted in Sec. 2, an explanation of the pre-seismic MHz and kHz EME silence is attempted in Sec. 3 and Sec. 4, respectively. In Sec. 5, the silence observed sequentially in all EM bands is interpreted from the percolation theory of fracture point of view. Finally, the results are discussed and summarized in Sec. 6.

**2. The up-to-date theoretical and laboratory background**
In this section, the information relative to the fracture-induced MHz-kHz EME that is available from the up-to-date published theoretical as well as laboratory works is synthetically presented. Based on the evidence highlighted in this section, an explanation of the silence feature observed in the MHz-kHz EM precursory signals is attempted in Secs 3, 4 and 5.

**2.1 Two stage model of EQ dynamics by means of MHz-kHz EM emissions**
Based on the concepts presented in the Introduction, we have studied the possible seismogenic origin of the MHz-kHz anomalies recorded prior to significant shallow EQs that occurred on land, trying to answer three fundamental questions: (i) *"How can we recognize an EM observation as a preseismic one?"* (ii) *"How can we link an individual EM precursor with a distinctive stage of the EQ preparation process?"* (iii) *"How can we identify precursory symptoms in EM observations that indicate that the occurrence of the EQ is unavoidable?"* The second question is the fundamental one. The answers to the other two questions should be rooted in the answer of the second one. EM phenomena associated with EQs may be recognized as real EQ precursors only when the physical mechanism of their origin is clarified (Pulinets and Boyarchuk, 2004; Cicerone et al., 2009).

The basic information that guided our thinking was the following well-established evidence. *An important feature, observed both at laboratory and geophysical scale, is that fracture-induced MHz radiation consistently precedes the kHz radiation, indicating that these two emissions correspond to different characteristic stages of the fracture / EQ preparation process* (Kumar and Misra, 2007; Qian et al., 1994; Eftaxias et al., 2002 and references therein; Baddari and Frolov, 2010). It is believed that the crack size varies in inverse proportion to the carrier frequency. Therefore, *the transition from the MHz to kHz EM emission signals the hierarchical development of failure, namely the transition from smaller to larger cracks.* As is consistent with this, laboratory experiments in the frame of mesomechanics support the concept that the MHz and kHz EM modes consecutively observed at the geophysical scale may signalize the transition of plastic flow localization from the meso-scale to the macro-scale, culminating in global fracture (Eftaxias et al., 2007). Our studies have verified that the MHz and kHz EM physical fields refer to completely different fracture mechanisms. More precisely, we have introduced the following two-stage model (Fig. 1) (Kapiris et al., 2004; Contoyiannis et al., 2005; Contoyiannis and Eftaxias, 2008; Papadimitriou et al., 2008; Eftaxias et al., 2007, 2010; Eftaxias, 2012 and references therein):

*(1) The initially emerging MHz EM field originates from the cracking in the highly heterogeneous material that surrounds the backbone of asperities, distributed along the stressed fault sustaining the system.*

*(2) The abrupt emergence of strong avalanche-like kHz EME is thought to be due to the fracture of the family of the asperities themselves.*

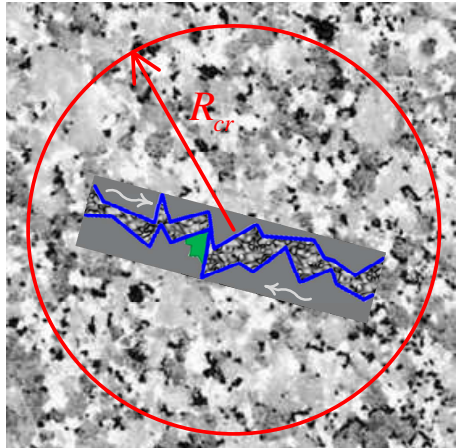

Fig. 1 A fault (blue lines) is embedded in an heterogeneous environment. The EQ preparation process at the first stage concerns an area of $R_{cr}$ radius around the fault where the heterogeneous media takes part in the fracture process, emitting MHz EME during the cracking of heterogeneous media. The symmetry breaking signals the transition from the phase of non-directional, almost symmetrical, cracking distribution to a directional localized cracking zone along the direction of the fault. The EQ is inevitable if and only when the asperities break (green highlighted area), emitting the kHz EME during the second stage, and then an EME silence follows.

### 2.1.1 Focus on MHz EM field

In analogy to the study of critical phase transitions in statistical physics, it has been argued that the fracture of heterogeneous materials could be viewed as a critical phenomenon either at the laboratory scale or at the geophysical scale (Chelidze, 1982; Allegre et al., 1982; Sornette and Sornette, 1990; Herrmann and Roux, 1990; Sornette and Sammis, 1995; Sornette and Andersen, 1998; Bowman et al., 1998; Kossobokov et al., 1999; Sornette, 2000; Moreno, 2000; Gluzman and Sornette, 2001; Guarino et al., 2002; Rundle et al., 2003 and references therein). We note that other researchers have associated brittle rupture with a first-order phase transition (Buchel and Sethna, 1997; Kun and Herrmann, 1999; Rundle et al., 2003 and references therein). However, this is still an open question. We have shown that the dynamics of the MHz EM emissions are characterized by antipersistency, namely, a negative feedback mechanism that "kicks" the crack-opening rate away from extremes, while this mechanism is rooted in the heterogeneity of the fractured system (Eftaxias et al., 2003, 2009; Kapiris et al., 2004; Contoyiannis et al., 2005). This evidence guided us to examine whether the observed MHz EM precursor can be described in analogy with a thermal, continuous, second-order transition in equilibrium. Recently, we have shown that this really happens (Contoyiannis et al., 2005). Moreover, combining ideas of Levy statistics, nonextensive Tsallis statistical mechanics, and criticality with features hidden in the precursory MHz time-series we have shown that a truncated Levy walk type mechanism can organize the heterogeneous system to criticality (Contoyiannis and Eftaxias, 2008). We underline that the MHz EM anomaly includes the crucial feature of *"symmetry breaking"* of critical phenomena. This feature permits the real-time, step-by-step monitoring of the damage evolution of the heterogeneous component in the focal area during mechanical loading. Indeed, our analysis reveals the following consecutive rupture process epochs as the EQ approaches (Contoyiannis et al., 2005; Eftaxias, 2012):

(i) The critical epoch (critical window) during which the short-range correlations between the opening cracks evolve to long-range ones.

(ii) The epoch of the "symmetry breaking" occurrence, namely, the transition from the phase of non-directional, almost symmetrical, cracking distribution to a directional localized cracking zone.

(iii) The completion of the "symmetry breaking". The rupture process has been obstructed at the boundary of the backbone of strong asperities. *We have proposed that the appearance of a MHz EM* *anomaly due to its nature is a necessary but not a sufficient requirement for the EQ occurrence. The completion of "symmetry breaking" simply means that the "siege" of asperities has already been started, i.e., the applied stresses have already been focused on them. The abrupt emergence of strong avalanche-like kHz EME activity reveals the fracture of asperities, if and when the local stress exceeds their fracture stress.*

The aforementioned results not only clearly discriminate the recorded MHz anomalies from the background EM noise, but are consistent with the endorsement of the MHz EM anomaly as reflection of an underlying fracture process of a heterogeneous medium.

**2.1.2 Focus on kHz EM field**
The notably crucial character of the suggestion that *the abrupt emergence of strong avalanche-like kHz EME activity reveals the fracture of asperities distributed along the fault sustaining the system, implying that the occurrence of the ensuing EQ is unavoidable as soon as kHz EME have been observed,* requires a strong support by well established fundamental arguments.

*Arguments in terms of statistical analysis.* First, based on a multidisciplinary statistical analysis we have shown that the kHz EM time series is characterized by the following crucial symptoms of an extreme phenomenon (Potirakis et al., 2011, 2012a,b,c; Eftaxias, 2012 and references therein): *(i) High organization or high information content.* Tools of information theory, concepts of entropy rooted in the extensive and nonextensive Tsallis statistical mechanics, and measures of complexity have been used in order to identify the aforementioned features (Shannon n-block entropy, Shannon n-block entropy per letter, conditional entropy, entropy of the source, Kolmogorov-Sinai entropy, T-entropy, Approximate Entropy –ApEn–, Fisher Information, Correlation Dimension, R/S analysis, Detrended Fluctuation Analysis, Fractal Dimension, and finally fractal wavelet spectral analysis). *(ii) Strong persistency*, indicating the presence of a positive feedback mechanism in the underlying fracto-EM mechanism that leads the systems out of equilibrium. *(iii) Absence of any footprint of a second-order transition in equilibrium or truncated-Levy walk type mechanism.* (iv) *Existence of clear preferred direction of fracture activities* (cf. Fig. 2) (Potirakis et al., 2012b). We note that according to the description of laboratory experiments and seismicity observations, AE and seismicity hypocenters tend to concentrate close to the nodal plane of a future main rupture (Mogi, 1968; Scholz, 1968; Ponomarev et al., 1997). *The aforementioned results are consistent with the endorsement of the kHz EM anomaly as reflection of an underlying extreme event.*

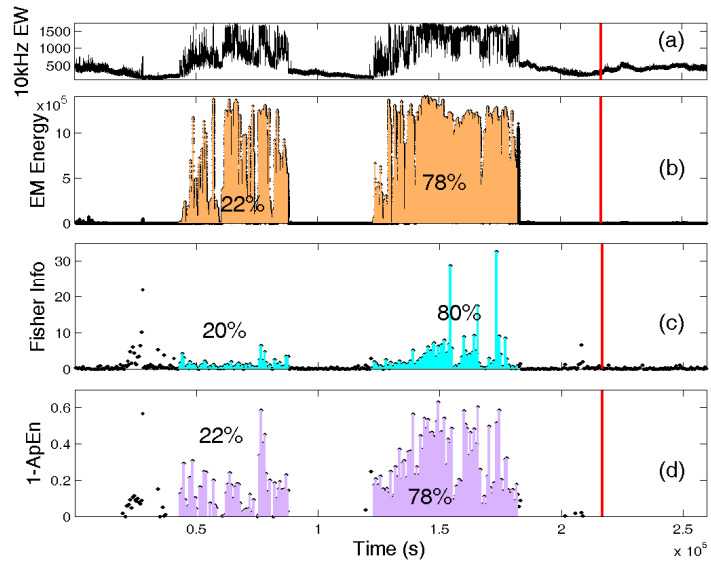

Fig. 2 (a) The two strong impulsive bursts in the tail of the recorded pre-seismic kHz EM emission prior to the Athens (Greece) EQ ($M = 5.9$) that occurred on September 7, 1999 (please refer to Fig. 1 in Papadimitriou et al. (2008)). For the specific signal excerpt, the EM Energy (b), the Fisher information (c) and approximate entropy (d) evolutions with time are presented. The colored areas indicate the energy, information and 1-ApEn corresponding to the two bursts. The first (left) burst is responsible for 22% of the EM energy, 20% of the Fisher information, and 22% of the ApEn, while the second (right) for 78%, 80%, and 78% respectively. All graphs are time aligned for direct reference. The time of the EQ occurrence is indicated by the thick vertical red line. (Fig 1. in Potirakis et al. (2012b)).

***Arguments in terms of universal structural patterns of the fracture process***. Nevertheless, from our point of view even the aforementioned multidisciplinary statistical analysis per se does not link the kHz EM phenomenon with the fracture of asperities. The above-noted statistical results are likely to offer *necessary but not sufficient criteria* in order to recognize a kHz EM anomaly as an indicator of the fracture of asperities. The crucial question was whether different approaches could provide additional information that would allow one to accept that the kHz EM anomalies signal the fracture of asperities. The basic information that guided our thinking was the following well-established evidence: Despite the complexity of the fracture process, there are universally holding scaling relations. The aspect of self-affine nature of faulting and fracture is widely documented (Mandelbrot, 1982; Huang and Turcotte, 1988; Turcotte, 1997; Sornette, 2000). From our point of view, universal structural patterns of the fracture process should be included in a precursor associated with the fracture of asperities.

We have shown that the above-mentioned requirement is satisfied. Indeed:
(i) Huang and Turcotte (1988) have stated that the statistics of regional seismicity could be merely a macroscopic reflection of the physical processes in EQ sources, namely, the activation of a single fault is a reduced self-affine image of regional seismicity. In this direction, we have shown that the populations of the EQs that precede a significant seismic event and occur around its epicentre, and the "fracto-EM EQs" that emerge during the fracture of the population of asperities follow the same statistics, namely, the relative cumulative number of EQs / EM-EQs vs. magnitude, either in terms of the traditional Gutenberg-Richter law (Kapiris et al., 2004) or by means of a model of EQ dynamics which is based on first principles of non-extensive statistical mechanics (Papadimitriou et al., 2008; Minadakis et al., 2012a, b). Furthermore, following the above-noted procedure we have further shown that the *activation of a single fault is a magnified self-affine image of laboratory seismicity as is observed by AE or EME* (Kapiris et al., 2004).
(ii) Fracture surfaces have been found to be self-affine following the persistent fractional Brownian motion (fBm) model over a wide range of length scales (Chakrabarti and Benguigui, 1998). It has been proved that the profile of the observed kHz EMEs follows this model (Contoyiannis et al., 2005; Minadakis et al., 2012a).
(iii) The roughness of fracture surfaces is quantitatively characterized by the Hurst exponent, $H$, since the average heights difference $\langle y(x) - y(x+L) \rangle$ between two points on a profile increases as a function of their separation, $L$, like $L^H$ with $H \sim 0.75$, weakly dependent on the nature of the material, on the failure mode, and on the spatial scale of fracture (Hansenand Schmittbuhl, 2003; Mourot et al., 2006). We have shown that the roughness of the profile of the observed kHz EM anomaly is consistent with the aforementioned universal $H-$value (Kapiris et al., 2004; Contoyiannis et al., 2005; Minadakis et al., 2012a).

**2.2 Are the AE and EME two sides of the same coin?**
The kHz EM emission is consistently launched in the tail of the fracture-induced EME both at the laboratory and the geophysical scale. Its absence during the EQ occurrence constitutes the most enigmatic feature in the study of EM precursors because older considerations had led to the erroneous conclusion that there is a contradiction between the laboratory and the field observed fracture-induced EME concerning the observed kHz EM silence. Indeed, the view up to a few years ago was that the observed AE and EME are two sides of the same coin. On the other hand, laboratory experiments in terms of AE were showing that this emission continues increasingly up to the time of the final collapse (Baddari and Frolov, 2010; Baddari et al., 2011; Schiavi et al., 2011). The combination of the above-mentioned information led to the erroneous conclusion that the appearance of kHz EM silence in field observations just before the EQ occurrence is a puzzling feature in terms of laboratory experiments.

However, recent accumulated laboratory evidence indicates that the recorded *AE and EME, in general, are not two sides of the same coin. There are two categories of AE signals, namely, AE signals which are associated with EME signals and AE signals which are not associated with EME signals* (e.g., Yamada et al., 1989; Rabinovitch et al., 1995; Morgounov, 2001; Mori et al., 1994, 2004a, b, 2006, 2009; Mori and Obata, 2008; Lacidogna et al. 2010; Baddari and Frolov, 2010; Baddari et al., 2011; Carpinteri, 2012). *Importantly, laboratory experiments performed recently reveal that the final stage of the failure process coincided in time with the maximum of AE and quiescence in EME*, while strong avalanche-like EME precedes this phase *(Morgounov, 2001; Baddari et al., 2011). Therefore, the kHz EM silence just before the final collapse is observed both at laboratory and geophysical scale.*

Baddari et al. (2011) emphasize that the alternating acoustic and EM quiescence in the kHz band was recorded during several different phases of block deformation and not only during the last phase of failure process; the great decrease of EME was accompanied by an increase of AE intensity and vice versa. This may indicate the alternation between two different mechanisms involved in a fracture process. The aforementioned laboratory results strongly support the hypothesis that in general EME and AE represent different phases of the destruction process and the combination of both emissions is important for the study of failure kinetics.

In the following we attempt to explain the reason which leads to the existence of the above-mentioned *two categories of AE signals*, *namely, the AE associated with EME and the AE not associated with EME*. We attempt this explanation in terms of: (i) the possible mechanism of generation of the observed fracture-induced EME; (ii) the dependence of the sensitivity of the emitted EME to the geometrical mode of the opening crack (shear or tensile), the brittleness index, Young modulus, and strength index of the material. We will see that the observed AE and EME are two sides of the same coin during the opening of tensile cracks in a material which is characterized by high brittleness index, Young modulus, and strength index. On the contrary, the phases of failure kinetics which are mainly characterized by frictional-noise-type rearrangements due to the frictional slip between the pre-formed fracture structures can be accompanied only by acoustic (EQ) events and not by EM events (Mori et al., 1994, 2004a, b; Mori and Obata, 2008). Subsequently, we try to tally correspond the two phases of the emerged strong kHz EME and the kHz EME silence to two distinct, last, phases of the EQ preparation process, namely to the fracture of entities of high index of strength, bulk modulus and brittleness via the opening of tensile cracks, on one hand, and the frictional-noise-type rearrangements due to the frictional slip between the pre-formed fracture structures which provide a bearing-like "lubrication" mechanism between fault plates, on the other hand.

### 2.2.1 On the origin of the fracture-induced EME

Several atomic models have been put forth to explain the origin of EM emission (Bahat et al., 2002 and references therein). Consider in particular the "movement of charged crack surfaces" (see Section II in Contoyiannis et al. (2005) and references therein). In this model, when a crack opens the rupture of interatomic bonds leads to intense charge separation (Langford et al., 1987; Cress et al., 1987). Direct evidence of surface charges of opposite polarity on fresh fracture surfaces has been provided by Wolibrandt et al. (1983). Simultaneous measurements of the electron, ion, and photon emission (fracto-emission) accompanying fracture support the hypothesis that charge separation accompanies the formation of fracture surfaces (Langford et al., 1987; Dickinson et al., 1998; Gonzalez and Pantano, 1990; Miura and Nakayama, 2000; Mizuno, 2002). On the faces of a newly created crack, the electric charges constitute an electric dipole or a multi-pole of higher order and due to the crack wall motion EM radiation is emitted (Miroshnichenko and Kuksenko, 1980; Khatiashvili, 1984; O' Keefe and Thiel, 1995; Molchanov and Hayakawa, 1995; Koktavy et al., 2004). Crack motion in fracture dynamics has recently been shown to be governed by a dynamical instability, which causes oscillations in the crack velocity and structure on the fracture surface. Evidence indicates that the instability mechanism is local branching, i.e., a multi-crack state is formed by repetitive, frustrated micro-fracturing events (Marder and Fineberg, 1996; Sharon and Fineberg, 1999). Laboratory experiments show intense EM fracto-emission during this unstable crack growth (Gonzalez and Pantano, 1990). In this unstable stage we regard the EME from the correlated population of opening cracks as a precursor of the final global instability (Contoyiannis et al., 2005).

If the above-mentioned interpretation is correct, the production of new fresh surfaces / rupture of bonds, due to cracking in the material, constitutes a *necessary* condition for the generation of EME. Mori and his colleagues (Mori et al., 1994, 2004a, b; Mori and Obata, 2008) emphasize that: the AE signals which are associated with EME signals are generated due to the creation of new fresh surfaces; on the contrary, AE signals which are not associated with EME signals are rooted in frictional noises due to the contact and/or the frictional slip/rolling between the pre-formed fracture surfaces.

### 2.2.2 On the sensitivity of the fracture-induced EM anomaly to the geometrical mode of opening cracks

Cracks can appear in solids under stress mainly in two different geometrical modes, tensile and shear. In the tensile mode, the displacements during the opening of cracks are perpendicular to the fracture propagation. On the contrary, in the shear mode the displacements are in the same direction as those of the crack propagation. Due to the fact that the tensile fracture mode leads to larger distances between the surface charges, compared to the shear mode, it has been proposed that an opening tensile crack behaves as a more efficient EM emitter (e.g., Yamada et al., 1989; Mori and Obata, 2008; Lacidogna et al., 2009, 2010, 2011; Baddari et al., 2011). Of course, the shear source EME should not be omitted; a certain separation between oppositely charged surfaces is guaranteed even in the shear fracture. However, the EME is more sensitive to the generation of tensile cracks than of shear cracks (Molchanov and Hayakawa, 1995).

### 2.2.3 On the sensitivity of the fracture-induced EM anomaly to elastic moduli, index of brittleness and index of strength

Elastic moduli, index of brittleness and index of strength constitute crucial parameters for the detection of AE or EME from the material experiencing "damage." It was found that an increase of Young modulus, index of brittleness, and strength enhances the AE or EME amplitude (Hudson et al., 1972; Gol'd et al., 1975; Nitsan, 1977; Khatiashvili, 1984; Carpinteri, 1986, 1989, 1990, 1994; Frid et al., 1999, 2000; Rabinovitch et al., 2002; Fukui et al., 2005). On the other hand, accumulated experimental and theoretical evidence indicate that the elastic modulus significantly decreases as damage increases, approaching to zero as the global fracture is approaching (Lockner and Maddenn 1991; Guéguen et al., 1997; Lin et al., 2004; Shen and Li, 2004; Amitrano and Helmstetter, 2006; Chen, 2012; Chelidze et al., 1988, 1990).

Characteristically, a very important class of models of material failure is the fiber bundle models (FBMs), which capture the most important aspects of material damage. In the frame of the discrete FBM, the stored elastic energy, $e_f(t)$, in a single fiber at the time of failure is given by the formula

$$e_f(t) = V_f E_0 \varepsilon_f^2(t) / 2 \quad (1)$$

where $V_f$ is the volume of the fiber, $E_0$ is the Young modulus of the fiber, and $\varepsilon_f(t)$ the strain in the fiber at time $t$ (Turcotte et al., 2003). Consequently, the elastic energy release should be greatly reduced during the softening mode. Laboratory studies show that EM energy release rate has a direct correlation with the elastic strain release rate (Kumar and Misra, 2007; Chauhan and Misra, 2008). Therefore, it is reasonable to assume that when an element fails *during the softening period* the emitted EM radiation is also greatly reduced. In summary, the sudden drop of stress just before the shock occurrence of its own accord signals a corresponding abrupt decrease in the amount of the elastic / fracto-EM energy that can be released, and finally provides a possible explanation of the observed EME gap.

### 2.3 Fracture induced EM anomalies in terms of percolation theory
In the following, we investigate the behavior of the elastic modulus of heterogeneous media during the fracture process and the way that this influences the emission rate of EM radiation in terms of percolation theory. Moreover, in the frame of this theory, we examine the discrimination of different thresholds of concentration of defects in a medium. The first of them could correspond to transfer phenomena and the subsequent two of them to different phases of fracture.

### 2.3.1 On the dependence of the fracture induced emission pattern on the elastic modulus
It has been shown that the problem of fracture of heterogeneous media corresponds mathematically to problems of the *percolation theory*, which describe quantitatively the connectivity of components in a non-homogeneous system (Chelidze, 1979, 1980a,b, 1982, 1986, 1987, 1993; Chelidze at al., 2006; Arbadi and Sahimi 1990). It appears that the elastic modulus, $M$, decreases from the initial value for an intact lattice over nearly five orders of magnitude according to the expression $M \propto M_0 (x - x_{cr})^{-3.6}$ where $x$ is the current concentration of breakages, and $x_{cr}$ is the percolation threshold for a given finite system; it has been found that $x_{cr} = 0.494$. Arbadi and Sahimi (1990) have studied models of mechanical breakdown in disordered solids by large-scale Monte-Carlo simulations. Their results also show that the Young's modulus $Y$ reduces to zero well before the total breaking of bonds, namely, when the fraction $p$ of the unbroken bonds is approximately $0.2$ (see Fig. 1 in Arbadi and Sahimi (1990)). A model of elastic wave emission and amplitude distribution during the failure has been developed (Chelidze and Kolesnikov, 1984; Chelidze 1993) which assumes that the emergence of a new defect is associated with an emission event and that the (conventional) emission amplitude $A_c$ generated by each elementary event depends directly on the increment of the size of the resulting defect cluster, induced by this single event,

$$A_c = A_0 \left( \left( \sum_{i=1}^{k+1} s_i \right)^2 - \sum_{i=1}^{k} s_i^2 \right)^{1/2} \quad (2)$$

where $k$ is the number of clusters linked by the elementary defects, $s_i$ is the number of sites in the $i-$th merging cluster, and $A_0$ is a conventional amplitude generated by the nucleation act of a single isolated defect. According to Eq. (2) the emission amplitude should increase with approach to the percolation threshold as merging of large clusters generates a larger dynamical event. Numerical experiments show that the amplitude spectrum of the quantity $A_c$ possesses basic characteristics of the AE of fracture, namely the increase in mean amplitude, and the growth of the amplitude dispersion as the percolation threshold is approached (Chelidze, 1986, 1993). On the other hand, in $A_0$ we take into account the current elastic modulus $M$ of the system, so that $A_0 = f(M)$, where $M \propto (p - p_c)^\alpha$, i.e., $M$ and correspondingly $A_0$ also change with increase of damage $p$; namely $M$ decreases according to the power law of Chelidze (1993); $p_c$ is the corresponding percolation threshold. Physical experiments involving artificial sequential damaging of the plastic lattice confirm the validity of the expression $M \propto (p - p_c)^\alpha$ with $a = -3.6$ (Chelidze et al., 1990).

A theoretical expression for energy emission, $\Delta E$, at the addition of one damaged site, expressed by means of fundamental percolation functions, is given in Chelidze (1986), where $\Delta E$ is calculated as a derivative of the mean size of finite clusters, $s$:

$$\Delta E = \frac{d}{dp} \sum_s s^2 n_s(p) \bigg/ \sum_s s n_s(p) = p\frac{ds(p)}{dp} + s(p) = \left( \sum_{i=1}^{k+1} s_i \right)^2 - \left( \sum_{i=1}^{k} s_i^2 \right) \quad (3)$$

where $n_s$ is the number of clusters of size $s$ and differentiating we take into account that the sum $\sum_s s n_s(p) = p$ for $p < p_c$.

For $p \to p_c$ we get:

$$\Delta E \approx |p - p_c|^{-(\gamma+1)} = s(1 + L_R) \quad (4)$$

where $L_R$ is the average number of cutting ("red", single) bonds in an incipient Infinite Cluster (IC) (Chelidze and Kolesnikov 1984). The same approach has been used by Sammis and Sornette (2002) and they obtained similar expressions for $\Delta E$. In the above treatment the elastic modulus $M$ of the system was considered as a constant, so it was actually ignored. This is true on the first stage of fracture process when the cracks form relatively small finite clusters, which can be considered as isolated soft inclusions in the (globally) rigid medium. However, when approaching the percolation threshold the elastic modulus depends on the concentration of defects $p$.

So the expression for emitted energy for $p \to p_c$ taking into account that $M$ is a function of $p$, can be written as:

$$\Delta E \approx \frac{dM}{dp}|p - p_c|^{-(\gamma+1)} \approx \alpha|p - p_c|^{\alpha-1}|p - p_c|^{-(\gamma+1)} \approx \alpha|p - p_c|^{\alpha-\gamma-2} \qquad (5)$$

Eq. (5) clarifies that the drastic decrease of elastic modulus in the last stage of fracture of an heterogeneous medium can significantly change the emission pattern described by Eq. (2), because the "standard amplitude" $A_0$ which includes the elastic characteristics of media, should also diminish when approaching the percolation threshold. Then, if the decrease of $A_0$ dominates over the increase of the second (right-hand) factor of Eq. (2) during the final stage of fracture, the mean amplitude of emission can even decrease with defect proliferation. This can cause apparent silence before the mechanical collapse.

### 2.3.2 On the discrimination of thresholds for transport and fracture in terms of percolation theory

The discrimination of thresholds for transport and fracture in terms of percolation theory also offers a quantitative explanation of the observed EM silence in various frequency bands. A percolation model, which considers the fracture of heterogeneous media as a critical phenomenon, was suggested in 1978 by T. Chelidze (first publications 1979, 1980a, b, 1982), and has been developed in many subsequent works (see, for example Herrmann and Roux, 1990). This model predicts the existence of "*critical points*" or *percolation transitions* in various properties of the material, when its characteristics change dramatically (Eftaxias and Chelidze, 2005). These transitions (metal-dielectric, gel-sol, permeable-impermeable, consolidated-unconsolidated) are observed for all the so-called generalized conductivities or transport-related properties of materials, which critically depend on the connectivity of transport channels. In many cases, the percolation thresholds for various properties are the same, which means that the corresponding transport processes have the same geometrical basis. *At the same time it is evident that the transition impermeable-permeable does not mean that the 3-dimensonal (3D) systems become mechanically unconsolidated.* The transition consolidated-unconsolidated, which allows the division of the 3D material into two parts, at least, demands much larger concentration of voids. Consequently, the percolation model of fracture is associated with two different thresholds of concentration of defects, $x$, namely:

(i) the *"hydraulic threshold"*, $x_c$, and

(ii) the *"mechanical or damage threshold "*, $x_m$, where the *Infinite Cluster (IC)* is formed, and the solid disintegrates (Chelidze, 1986).

As a rule, the relation $x_m > x_c$ holds.

The EQ is considered as a shear displacement along the fault plane. However, a layer of thickness $h$ of the material should contain the *IC*. Chelidze (1986) has proposed that the associated problem of shear displacement along the fault can be formulated in terms of mechanical percolation as following: "find the concentration $x_{mf}$, at which the *Flat Infinite Cluster (FIC)* of voids (cracks) spanning the layer of thickness $h$ would be formed in a 3D body." The problem of such an *FIC* has been solved in the percolation theory (Shklovsky, 1984; Chelidze, 1987). The required critical concentration for the formation of *FIC*, $x_{mf}$, is (Chelidze 1986):

$$x_{mf} = x_m \left[1 + d(r_s / h)^{1/\nu_3}\right] > x_m \qquad (6)$$

where $r_s$ is the mean distance between defects, $d$ is a constant close to unity, and $\nu_3 = 0.9$ for 3D. The *FIC* mechanical percolation model is very similar to the well-known physical models of seismic process evolution, namely, dilatancy-diffusion and avalanche fracturing models (Mjachkin et al., 1975; Scholz et al, 1973). Based on the above-mentioned considerations, it might be concluded that many transport properties are activated before the collapse of the sample (Paterson, 1978; Chelidze, 1986). We note that the hydraulic percolation thresholds correspond to a critical crack density of about 0.1, while experimental data show that the mechanical threshold is of the order of 1 (Zhang et al., 1994a, b).

### 3. An explanation of the observed pre-seismic MHz EME silence
In the following, we attempt an explanation of the observed pre-seismic MHz EME silence in terms of the two-stage model, laboratory experimental evidence, and percolation theory as presented in Sec. 2, as well as in terms of the foreshock activity.

### 3.1 Arguments by means of the two-stage model of EQ generation
The consistently observed pre-seismic silence of the MHz EME is fully understood in the frame of the above-mentioned two-stage model of EQ generation (see SubSec. 2.1): this emission stops well before the EQ occurrence, namely when the fracture of the heterogeneous material that surrounds the backbone of asperities distributed along the stressed fault sustaining the system has been completed and the stresses built up on the family of asperities; the preparing EQ will occur if and when the local stress exceeds the fracture stresses of these strong entities. Accumulated experimental and theoretical evidence support this proposal, as follows.

### 3.2 Arguments by means of laboratory experiments
Laboratory experiments by means of AE and EME strongly justify the appearance of MHz silence prior to the global failure; the MHz activities are rooted in earlier lower levels of fracture in comparison to that of the kHz activity (Eftaxias et al., 2002 and references therein; Baddari and Frolov, 2010). Experimental studies by means of mesomechanics (see SubSec. 2.1) support the consideration that the MHz activities are launched earlier in comparison to the kHz activities (Eftaxias et al., 2007).

### 3.3 Arguments by means of foreshock activity
We recall that the precursory MHz EM phenomenon can be described in terms of a second order thermal phase transition in equilibrium. It is reasonable to accept that such a phenomenon should be associated with a stage of the EQ preparation process which precedes the final stage of the fracture where the *extreme* seismic event is formed; it cannot originate from the final stage. Consequently it might be deduced that the observed pre-seismic MHz EM silence before the seismic shock occurrence reflects a well-established precursory feature of the fracture process. However, the foreshock activity and the precursory fracture-induced MHz EM emission constitute two sides of the same coin. A question which logically arises is whether the foreshock activity during the last week before the main shock occurrence also behaves as a critical phenomenon. A recent analysis leads to a positive answer. Based on the recently introduced concept of the *natural time* (Varotsos et al., 2011), we have proved that (Potirakis et al., 2013): (i) the foreshock seismic activity that occurs in the region around the epicenter of the oncoming significant shock a few days up to one week before the main shock occurrence, and (ii) the observed MHz EM precursor which emerges during the same period, both behave as critical phenomena. This result strongly supports the seismogenic origin of the observed EM precursor, its behavior as a critical phenomenon, and finally the hypothesis that the observed MHz EM silence before the global instability occurrence is an expected and non-puzzling feature in the frame of fracture process.

In the following paragraph we attempt to explain the observed MHz EM silence based on the pioneering work of Chelidze, who first studied the EQ preparation process by means of the percolation theory (Stauffer, 1985).

### 3.4 Arguments by means of percolation theory
In paragraph 2.3.1 it has been shown by means of fundamental percolation functions that the extreme decrease of elastic modulus when the percolation threshold is approaching diminishes the fractured-induced AE or EME causing apparent silence before the mechanical collapse of the heterogeneous medium. We recall that laboratory experiments have shown that an increase of the elastic modulus enhances the AE or EME amplitude (see paragraph 2.2.3), while accumulated experimental and theoretical evidence indicate that the elastic modulus significantly decreases as damage increases, approaching to zero as the global fracture is approaching (see paragraph 2.2.3).

*Based on this evidence, the appearance of silence in the MHz EM band before the EQ occurrence is an expected feature according to mechanical percolation theory.*

### 4. An explanation of the observed pre-seismic kHz EME silence

Guided by the laboratory results presented in SubSec. 2.2 that: (i) there are two categories of AE signals; (ii) an opening tensile crack behaves as a more efficient EM emitter than a shear crack; (iii) the higher the brittleness index, strength index, and elastic modulus the stronger is the emitted EME, we propose that the consecutively emerging strong impulsive kHz EME and kHz EM silence represent two different distinct phases of the destruction process: the fracture of the asperities and the following frictional noise type mechanism phase that includes the occurrence of the main seismic event, respectively.

Note that laboratory studies support the aforementioned two-phase proposal. Characteristically, Zhang et al. (2009) have recently presented a physical theory for brittle failure that aims to explain both the phenomenological and micro-structural observations. Based on experimental results, the authors conclude that "the localized failure process of rock experiences two stages: the *brittle* breakage stage *(bond rupture)* and the sliding stage *(frictional resistance of failure plane mobilization).*"

**4.1 Focus on the first phase**
In SubSec. 2.1 we have reported accumulated evidence supporting the proposal that the observed kHz EME is rooted in the fracture of multi-asperity system distributed along the activated fault. The fracture of this system obeys all the requirements to form a population of efficient kHz EM radiations as they have been presented in SubSec. 2.2. More precisely, the asperities are characterized by high index of brittleness, high index of strength and high bulk modulus. We do expect, on mechanical grounds, that large EQs will nucleate on strong entities, which can store sufficient stress to produce a rupture that will not be easily stopped (Sornette, 1999 and references therein). On the other hand, the fracture of asperities happens through the opening of tensile cracks. Laboratory studies clearly suggest that the formation of the macroscopic fracture plane is more likely to be the result of a tensile than a shear process (Hallbauer et al., 1973).

Herein we present more evidence associating the observed pre-seismic kHz EME with the fracture of asperities. From experimental studies on the frictional movement between rock surfaces, it is known that when the static friction resistance is overcome the frictional surfaces suddenly slip, lock and then slip again in a repetitive manner (Matcharashvili et al., 2011). If one asperity fails, the stress is transferred to the adjacent element on which an induced failure could occur. The unstable transition from the static friction to kinetic one is known as "stick-slip" state; EQs have long been recognized as resulting from a stick–slip frictional instability (Scholz, 1998). Recent works reveal that "stick-slip" friction events observed in the laboratory and EQs in continental settings, even with large magnitudes, have similar rupture mechanisms (Lockner and Okubo, 1983; McGarr and Fletcher, 2003; McGarr et al., 2010). During the fracture of an asperity a kHz EM avalanche, a "fracto-EM EQ" is emitted. The avalanche-like feature of the observed kHz EME activity is consistent with the suggestion that the population of the observed "fracto-EM EQs" reveal the population of the fractured asperities. We note that the time scale of the final tertiary stage of creep is characterized by an *avalanche* in the strain rate (Morgounov, 2001), as well.

The fracture of a strong asperity should be *accompanied by a sharp drop in stress.* Indeed, laboratory studies suggest that the strong avalanche-like *EM* signals are generated only during sharp drops in stress, while the amplitude of the emitted EM fields is proportional to the stress rate; (Fukui, 2005; Carpinteri et al., 2011, 2012; Lacidogna et al., 2011). These sharp drops are attributed to a rapid decay of the mechanical properties, generated by formation of new micro-cracks during the loading process (Carpinteri et al., 2012).

We reiterate that the preseismic kHz EME behaves as a fractal radiation (see SubSec. 2.1.2). Muto et al. (2007) conducted a friction laboratory experiment simulating the motion of an asperity on a fault plane, and observed photon emissions due to electric discharges by dielectric breakdown of ambient gases from friction contact between rock minerals. Those authors conclude that the frictional discharge occurring at asperities on the fault plane can be one of the origins of fractal EM radiation observed prior to EQs.

**4.2 On the second phase (EM silence)**
The herein examined crucial question is whether, within the frame of materials science, mechanisms have been proposed which could explain the absence of kHz EM radiation just before the EQ occurrence. We argue that such mechanisms have been proposed, while recent laboratory and numerical experiments fully justify the hypothesis that the emerged silence is the last precursory feature of an imminent significant EQ and not a puzzling feature.

**4.2.1 From microscopic to macroscopic mechanics**
We propose that the resolution of the enigmatic feature under study may be hidden in the resolution of another well-known paradox of EQ dynamics which the geophysicists call *the heat-flow paradox* (Anoosherhpoor and Brune, 1994; Alonso-Marroquin et al., 2006 and references therein). According to our common sense, when two blocks grind against one another, there should be friction, and that should produce heat. However, measurements of heat flow during EQs are unable to detect the amount of heat predicted by simple frictional models. Calculations using the value of rock friction measured in the laboratory, i.e., a typical friction coefficient between 0.6 and 0.9, lead to overestimation of the heat flux. As an example, one refers in this context to the heat flow observations made around the San Andreas fault, which show that the effective friction coefficient must be around 0.2 or even less (Alonso-Marroquin et al., 2006 and references therein).

Various relevant mechanisms have been proposed, however, the correct assessment of frictional heat production during shear is a central issue. There are different natural systems which fail through a localized failure in narrow shear zones. These so-called shear bands appear, for example, in granular packings (Astrom et al., 2000). Tectonic faults are a characteristic example of shear failure in narrow zones; many naturally occurring faults and shear zones contain regions of granular material. The gouge particles, i.e., fragmented rock inside the fault zone, play a fundamental role in influencing the macroscopic behavior of these systems (Latham et al., 2006; Welker and McNamara, 2011). The formation of rolling bearings inside the gouge has been introduced as a possible explanation for a substantial reduction of the effective friction coefficient. A simplified picture assumes that the gouge is filled with more or less round grains which, as the plates move, can roll on each other thus reducing the amount of frictional dissipation. Granular dynamics simulations (Astrom, et al., 2000; Baram and Herrmann, 2004; Herrmann, et al., 2004) have demonstrated the *spontaneous* formation of such bearings, while laboratory experiments also demonstrated the spontaneous formation of such bearings (Veje et al., 1999). We note that Wilson et al. (2005) have shown that gouge from the San Andreas fault, California, with ~160 km slip, and the rupture zone of a recent EQ in a South African mine with only ~0.4 m slip, display similar characteristics, in that ultrafine grains approach the nanometer scale, gouge surface areas approach $80 \mathrm{m}^2 \mathrm{g}^{-1}$, and grain size distribution is non-fractal. Wilson et al. (2005) also propose that the observed fine-grain gouge is not related to quasi-static cumulative slip, but is instead formed by dynamic rock pulverization during the propagation of a single EQ.

The above-mentioned picture which refers to the phase that follows the fracture of asperities, seems to justify the observed EM silence. As the plates move, the grains can roll on each other as rolling bearings and thus the resulting deformation is not accompanied by significant fracturing, i.e., by formation of new fracture surfaces via the opening of tensile cracks. Such frictional-noise-type rearrangements due to the frictional slip between the pre-formed fracture structures can be accompanied only by acoustic (corresponding to EQ in geophysical scale) events and not by EM events (Mori et al., 1994, 2004a, b; Mori and Obata, 2008). As already noted in SubSec. 2.2, laboratory experiments reveal that AE of maximal intensity are recorded directly before and during the fragmentation of the specimen, while strong avalanche-like EME precedes this phase (Baddari et al., 2011; Morgounov, 2001).

Numerical studies support the proposal that the final stage of the failure process is not accompanied by significant fracturing. Park and Song (2013) have presented a new numerical method for the determination of contact areas of a rock joint under normal and shear loads. They report that "after the peak stage, the contact area ratio decreased rapidly with increasing shear displacement, and few inactive elements came into contact until the residual stage. At the residual stage, only small fractions of 0.3% were involved in the contact."

As already noted, the efficiency of generating EME is thought to be smaller in shear cracks than in tensile cracks. Yamada et al. (1989) conclude that "the final main shock is considered to be a result of shear faulting, which may occur by connecting numerous cracks produced previously", while Stanchits et al. (2006) conclude that "Close to failure the relative contribution of tensile events decreased. A concomitant increase of double-couple events including shear, suggests that shear cracks connect previously formed tensile cracks." Therefore, it is reasonable to expect that the anomalous EME will be observed only before an EQ but not during the main shock (Yamada et al., 1989). According to Benioff (1951), creep may be purely compressional, purely shearing, or a combination of two. In the last case, the compressional phase occurs first. Importantly, the author proposes that the shearing phase follows an interval of 0.01 to 2.4 days. This time-window is in agreement with the observed duration of the kHz EM gap.

### 5. Interpretation of the sequential appearance of the ULF-MHz-kHz precursors in terms of percolation theory of fracture

It must be mentioned here that well-established DC-Ultra Low Frequency (ULF) EM precursors (<1Hz) have been observed to appear earlier than the MHz and kHz ones. Their lead time extends up to three months before the EQ occurrence (Varotsos, 2005; Molchanov and Hayakawa, 2008). In this section we try to explain why the DC-ULF, MHz, and kHz EM anomalies appear one after the other and thus why the DC-ULF EM precursors are also absent during the EQ occurrence. The relevant arguments are based on the discrimination of thresholds for transport and fracture in terms of percolation theory (see paragraph 2.3.2).

It is reasonable to assume that the various short-lived EM preseismic anomalies should be observed with the following order of time: (i) First, precursory phenomena associated with *transport properties* appear well before the collapse. (ii) Second, fracture-induced precursory phenomena associated with the *formation of the IC* are launched. The MHz EM activity is such a phenomenon. It might be considered that the precursory MHz EME, which is emitted during the last week before the shock occurrence, is associated with the mechanical threshold $x_m$, namely, with the formation of flat cluster of thickness $h$. During this formation, the fracture or grinding of weak heterogeneous material distributed around the main fault emits the antipersistent MHz radiation. Its behavior as a second-order phase transition as well as the observed *symmetry breaking* during the temporal evolution of this radiation strongly support the aforementioned hypothesis, especially the association of the MHz precursor with the formation of flat cluster. (iii) Third, the fracture-induced precursory phenomena rooted in the *formation of the FIC* emerge. The kHz EME is such a phenomenon. The observed persistent kHz EM activity from a few decades of hours up to a few minutes before the EQ occurrence might emerge around the critical mechanical threshold $x_{mf}$, i.e., during the fracture of high strength and brittle asperities sustaining the system.

Different mechanisms for the generation of the first appearing DC-ULF precursors have been proposed, like the ones presented by Varotsos (2005), and Molchanov and Hayakawa (1995), while publications have also suggested that electrokinetic phenomena can provide the basis for the generation of ULF electrical and magnetic precursors (Mizutani et al., 1976; Surkov et al., 2002 and references therein).

Here we focus on the electrokinetic phenomena. Electrokinetic electric and magnetic fields result from fluid flow through the crust in the presence of an electric double layer that is formed at the solid-liquid interface. This electric double layer is made up of a layer of ions (the Helmholtz layer) anchored to the surface of the solid (e.g., the rock) and as a diffuse mobile layer of ions of opposite sign (the Gouy-Chapman zone) which extends into the liquid phase. Fluid flow in this system transports the ions in the fluid in the direction of flow, and thus electric currents result. In this way, when a fluid flows through a porous medium, a potential will be developed across the length of the sample, because of the relative motion between the solid and the liquid. Thus, it has been suggested that a fracture-induced precursory ULF streaming potential appears. *The electrokinetic phenomenon, as a mechanism of generation of the ULF pre-seismic signals, should be associated with the "hydraulic threshold" within the frame of the percolation model, because it is not observed in partially saturated disperse systems.* A recent work (Surkov et al., 2002) strongly supports the above-mentioned hypothesis. Electrokinetic effects possibly associated with the EQ preparation process have been theoretically investigated, under the assumption that the structure of pore space around a hypocentral zone has fractal properties. This study verifies well-established experimental results, namely, the predicted dependence of the electrical anomalies amplitude $E$ on the EQ magnitude, the weak values of magnetic variations at far distances, and the presence of highly conductive channel(s) between the source and observation in order for the precursory electrical anomalies to be detectable. We note that, as the electrokinetic processes are relatively slow, their contribution to higher frequency bands is negligible; the electrokinetic phenomena decay at high frequencies. On the other hand, direct EM emissions at high frequencies are not detectable at this stage of fracture, as the concentration of cracks at hydraulic threshold is very small, while their orientation is random and the emitted EM fields cancel each other.

## 6. Discussion- Conclusions

In this work we studied a paradox associated with a series of different EM seismic precursors ranging from DC to MHz frequency bands, namely, their observed silence before the main seismic shock occurrence. We conclude that the considered puzzling features of the MHz-kHz fracture-induced EME are well-justified features and not paradoxical ones if we employ for their resolution a two-stage model for the EQ generation process and the percolation theory of fracture. Moreover, the silence of DC-ULF precursors is also explained based on percolation theory and the hypothesis that they are generated due to the electrokinetic phenomenon. Well- established concepts of the rupture process, in terms of: universal structural patterns of the fracture process, recent laboratory experiments which prove that AE and EME, in general, are not the two sides of the same coin, numerical and theoretical studies of fracture dynamics, the behavior of the elastic modulus of heterogeneous media during the fracture process, the discrimination of different thresholds of concentration of defects in a medium, critical phenomena, and micromechanics of granular materials, support the above conclusion. We underline that the EM silence before an EQ is not a puzzling but an expected feature.

In this contribution, we have based our analysis on the two-stage model for fracture-induced EME presented in Sec. 2 and we have presented much strong evidence that supports the specific hypothesis. However, it should be noted that other hypotheses for the origin of the observed pre-EQ EME have also been expressed that do not agree with the hypothesis of fracture-induced EME. For example, it has been proposed that pre-EQ EME originate from discharges in the atmosphere modified by the pre-EQ effects (e.g., Sorokin et al., 2011). Which hypothesis is true will be proved in the future, as the corresponding research evolves.

A crucial question arises immediately after the occurrence of a significant EQ referring to whether the EQ that occurred was the main shock or a foreshock of an ensuing larger EQ.

AE and EME studies of cracking in rocks have demonstrated that, in general, during cyclic loading, the level of AE and EME increases significantly when the stress exceeds the maximum previously reached stress level (Khatiashvili, 1984; Chelidze, 1986 and references therein; Li and Nordlund, 1993; Lavrov, 2005 and references therein; Mori and Obata, 2008; Mavromatou et al., 2008; Shkuratnik and Lavrov, 1996). This phenomenon was first reported in metals (Kaiser, 1953) and is now known as the Kaiser "stress-memory" effect. If we accept that the EME Kaiser effect is extended to the geological scale, then we suggest the following hypothesis should be thoroughly investigated in the future:
An observed continuous absence of fracture-induced MHz-kHz EME after the occurrence of a significant EQ implies that the launched significant EQ was the main shock. If a new MHz-kHz EME sequence emerges during a time-period of a few tens of hours after the already occurred significant EQ, this means that a new significant EQ is preparing, however in a new undamaged region of the Earth's crust out of the fractured region of the already occurred EQ.